\documentclass[prd,aps,nofootinbib,floatfix,11pt]{revtex4}
\usepackage{amsmath,graphicx,epsfig,amssymb,dsfont,mathtools,}
\usepackage[usenames]{color}
\usepackage{ulem} 
\usepackage{bigstrut}
\usepackage{slashed}
\usepackage{multirow}
\usepackage{subfigure}

\allowdisplaybreaks

\begin{document}
\title{Weak radiative decay {\boldmath$\Lambda_{c}^{+}\to\Sigma^{+}\gamma$}
  using light-cone sum rules}

\author{Yu-Ji Shi$^{1}$~\footnote{Email:shiyuji92@126.com},
  Zhi-Peng Xing~$^{2}$~\footnote{Email:zpxing@sjtu.edu.cn (corresponding author)}
  and Ulf-G. Mei{\ss}ner~$^{1,3,4}$~\footnote{Email:meissner@hiskp.uni-bonn.de}}
\affiliation{$^1$ Helmholtz-Institut f\"ur Strahlen- und Kernphysik and Bethe Center 
  for Theoretical Physics,\\ Universit\"at Bonn, 53115 Bonn, Germany\\
  $^{2}$ Tsung-Dao Lee Institute, Shanghai Jiao Tong University, Shanghai 200240, China\\
  $^3$ Institute for Advanced Simulation, Institut f{\"u}r Kernphysik and J\"ulich Center for Hadron Physics,
  Forschungszentrum J{\"u}lich, D-52425 J{\"u}lich, Germany\\
$^4$ Tbilisi State University, 0186 Tbilisi, Georgia}

\begin{abstract}
We calculate the decay width of the $\Lambda_{c}^{+}\to\Sigma^{+}\gamma$ using light-cone sum rules.
For the initial quark radiation an effective Hamiltonian is constructed, where the internal quark
line shrinks to a point. The final quark radiation is studied within the full theory. The leading
twist light-cone distribution amplitudes of the $\Sigma^+$ serve as the non-perturbative input for
the sum rules calculation, and the perturbative kernel is calculated at  leading order. The
branching fraction we obtain is ${\cal B}(\Lambda_{c}^{+}\to\Sigma^{+}\gamma)=1.03\pm 0.36 \times 10^{-4}$,
which is below the recent upper limit $<2.6 \times 10^{-4}$ given by the Belle collaboration.
\end{abstract}
\maketitle

\section{Introduction}
Weak radiative decays of charmed hadrons are an ideal platform for investigating the interplay
of the strong and the weak interactions.  Unlike the flavor-changing neutral-current transition of
bottom hadrons, the penguin contribution in such charm decays is highly suppressed. As a result,
the weak radiative decay of charmed hadrons are Cabibbo-favored and dominated by long-distance
non-perturbative effects, where the decay is induced by internal W-exchange bremsstrahlung processes
such as $cd\to us\gamma$. Studying the weak radiative decays of charmed hadrons both from the
experimental and the theoretical side can help us to understand the strong dynamics inside hadrons. 

Over the past few decades, there are several measurements of the weak radiative decays of charmed meson
\cite{Belle:2003vsx,BaBar:2008kjd,Belle:2016mtj}, and the corresponding theoretical researches
\cite{Cheng:2010ry,Artuso:2008vf,Fajfer:2006zqn,Fajfer:2005ke,Pakvasa:2005pp,deBoer:2018zhz,Dias:2017nwd,deBoer:2017rzd,Adolph:2022ujd,Adolph:2021ncg,Gisbert:2020vjx,Fu:2018yin,Biswas:2017eyn}. However, the experimental
researches in the charmed baryon sector are rare. Recently, the Belle collaboration announced the first
search for the weak radiative decays $\Lambda_{c}^{+}\to\Sigma^{+}\gamma$ and $\Xi_c^0\to \Xi^0 \gamma$
\cite{Belle:2022raw}, where the upper limits for their absolute branching fractions are given as:
\begin{align}
{\cal B}_{\rm exp}(\Lambda_{c}^{+}\to\Sigma^{+}\gamma) <2.6 \times 10^{-4},~~~~
{\cal B}_{\rm exp}(\Xi_c^0\to \Xi^0 \gamma) <1.7 \times 10^{-4}. \label{eq:upperlimit}
\end{align}
On the theoretical side, the corresponding branching fractions  have been predicted by various
theoretical approaches, which include a modified nonrelativistic quark model \cite{Kamal:1983zt},
the constituent quark model \cite{Uppal:1992cc} and the effective Hamiltonian approach combined
with the pole model \cite{Cheng:1994kp}. The theoretical predictions of the branching fractions
of $\Lambda_{c}^{+}\to\Sigma^{+}\gamma$ and $\Xi_c^0\to \Xi^0 \gamma$ are in the range
$(4.5-29.1)\times 10^{-5}$ and  $(3.0-19.5)\times 10^{-5}$, respectively. Most of these predictions
are consistent with the experimental constraints given above, while the one from the constituent
quark model are slightly larger than the upper limits in Eq.~(\ref{eq:upperlimit}). 

Nowadays, except the model-based theoretical approaches mentioned above, there is no model-independent
calculation for the weak radiative decays of charmed baryons. In this work, we will  fill this gap and calculate the
decay width of the  $\Lambda_{c}^{+}\to\Sigma^{+}\gamma$ with the use of  light-cone sum rules (LCSR).
In terms of the initial quark radiation, following Refs.~\cite{Uppal:1992cc,Cheng:1994kp}, we 
construct an effective Hamiltonian to simplify the calculation.  Since the radiating quark comes
from the heavy baryon $\Lambda_c^+$, its velocity can be assumed to be parallel to the velocity of the
$\Lambda_c^+$. This enables us to shrink the internal off-shell quark line to a point so that
the decay amplitude can be effectively induced by a local Hamiltonian of $cd\to us\gamma$. In terms
of the final quark radiation, since the final state $\Sigma^+$ is a light baryon, thus  we cannot make the
same assumption on its composite quark velocities. Therefore we have to treat the final quark
radiation in the full theory. The leading twist light-cone distribution amplitudes (LCDAs) of
the $\Sigma^+$ will serve as the non-perturbative input for the sum rules calculation. These
LCDAs are taken from the latest Lattice QCD calculation with $N_f=2+1$ \cite{RQCD:2019hps}. Furthermore, the perturbative kernel will be calculated at leading order. It should be mentioned 
that these LCDAs are defined according to the light-cone expansion, which are most reliable when the quark masses vanish. Therefore the LCDAs of 
$\Xi^0$ are not as good as those of $\Sigma^+$ which has less massive $s$ quarks, and we will not consider $\Xi_c^0\to \Xi^0 \gamma$ in this work.

This paper is organized as follows. In Sec.~\ref{sec:decayAmps}, we construct an
effective Hamiltonian for the initial quark radiation in the $\Lambda_{c}^{+}\to\Sigma^{+}\gamma$ decay
and express the decay amplitude by several calculable matrix elements. In Sec.~\ref{sec:LCSRatHadronlevel},
we define suitable correlation functions to calculate the decay amplitude at the hadron level.
In Sec.~\ref{sec:LCSRatQCDlevel}, we perform the QCD level calculation for the correlation function
defined above with the use of $\Sigma^+$ LCDAs. Sec.\ref{sec:numericalResult} contains the numerical
results on the decay amplitudes and branching fraction.  We will also compare them with those
from literature. Sec.~\ref{sec:conclusion} is a brief summary of this work.

\section{Decay amplitudes for initial and final radiation}
\label{sec:decayAmps}
\begin{figure}
\begin{center}
\includegraphics[width=0.8\columnwidth]{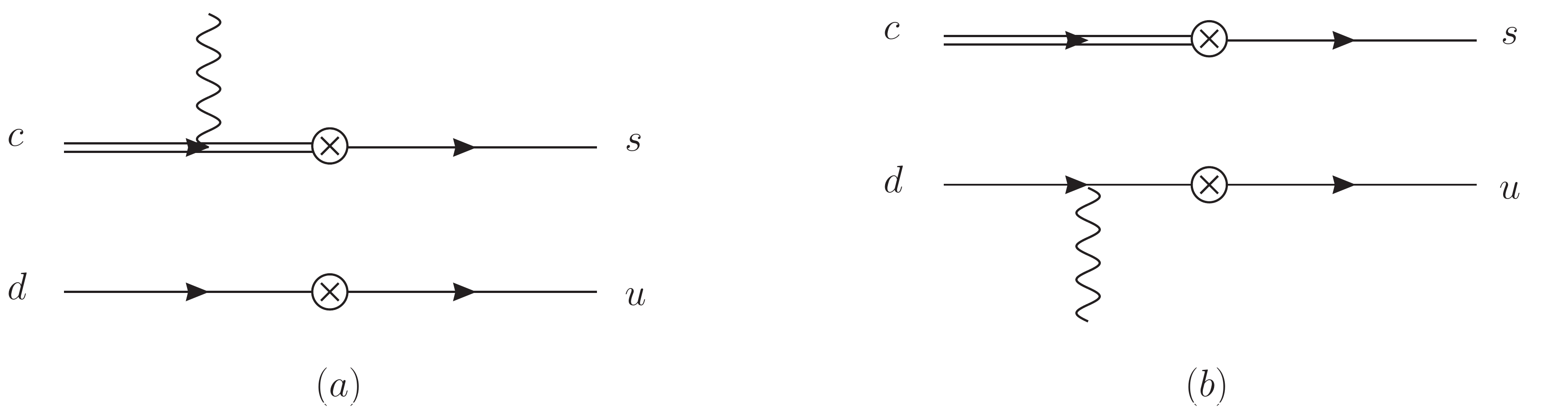} 
\caption{W-exchange bremsstrahlung processes $cd\to us\gamma$ induced by ${\cal O}_1$, where the
  double crossed dots denote ${\cal O}_{1,2}$. The diagrams for ${\cal O}_2$ are similar, just
  exchanging $u$ and $s$. }
\label{fig:EffectiveH} 
\end{center}
\end{figure}

The weak effective Hamiltonian contributing to the $\Lambda_{c}^{+}\to\Sigma^{+}\gamma$ decay reads
\begin{align}
{\cal H}_{\rm eff}&=\frac{G_F}{\sqrt 2}V_{cd}V^{*}_{ud}\left(C_1 {\cal O}_1+C_2 {\cal O}_2\right),\nonumber\\
{\cal O}_1&=\bar s \gamma^{\mu} (1-\gamma_5)c\  \bar u \gamma_{\mu} (1-\gamma_5)d,\nonumber\\
{\cal O}_2&=\bar u \gamma^{\mu} (1-\gamma_5)c\  \bar s \gamma_{\mu} (1-\gamma_5)d,
\end{align}
where the $C_{1,2}$ are the Wilson coefficients. Fig.~\ref{fig:EffectiveH} shows the W-exchange bremsstrahlung
processes $cd\to us\gamma$. In the case of initial radiation where the photon is emitted by the
$c$ or $d$ quark as shown in Fig.~\ref{fig:EffectiveH}, following the approach given in
Ref.~\cite{Cheng:1994kp} we can construct an effective Hamiltonian to simplify the calculation.
Here, we take the $c$ quark radiation in the ${\cal O}_1$ contribution, namely Fig.~1(a) as an example
to illustrate the procedure. 

The amplitude of Fig.\ref{fig:EffectiveH}(a) reads 
\begin{align}
{\cal A}_{{\rm Initial},c}^{{\cal O}_{1}}=i\frac{G_{F}}{\sqrt{2}}V_{cs}V_{ud}^{*}\varepsilon^{*\mu}(k)
\bar{s}(p_{s})\gamma^{\nu}(1-\gamma_{5})\frac{\slashed p_{c}-\slashed k+\bar m_{c}}{(p_{c}-k)^{2}-
  \bar m_{c}^{2}}\gamma_{\mu}c(p_{c})\bar{u}(p_{u})\gamma_{\nu}(1-\gamma_{5})d(p_{d}),
\label{eq:ampO1c}
\end{align}
where $p_{c,s,u,d}$ are the on-shell quark momenta, $\bar m_{c,d}$ are the constituent quark masses
in the  $\Lambda_{c}$ and $k$ satisfies $k^2=0$ and $k\cdot \varepsilon=0$.
Since the initial $c,\ d$ quarks are confined in the  heavy baryon $\Lambda_{c}$, we can assume
that $c$, $d$ and $\Lambda_{c}$ have the same velocity, in other words $p_{c,d}=(\bar m_{c,d}/m_{\Lambda_c})
p_{\Lambda_c}$. Thus the denominator of Eq.~(\ref{eq:ampO1c}) becomes
\begin{align}
(p_{c}-k)^{2}-m_{c}^{2}=\frac{\bar m_c}{m_{\Lambda_c}}(m_{\Sigma}^2-m_{\Lambda_c}^2),
\end{align}
which implies that effectively the internal off-shell quark line shrinks to a point. Further,
the numerator of Eq.~(\ref{eq:ampO1c}) can be simplified by using the equation of motion of the
$c$ quark. For the case of $d$ quark radiation the derivation is almost the same. Finally,
the amplitude in Eq.~(\ref{eq:ampO1c}) can be effectively generated by the following Hamiltonian
\begin{align}
{\cal H}_{{\rm eff}}^{{\cal O}_{1}} & =\frac{G_{F}}{\sqrt{2}}V_{cs}V_{ud}^{*} \ C_1 \sum_q
\left[A_{\mu}J_{{\cal O}_{1},q}^{\mu}-\frac{i}{2}F_{\mu\nu}K_{{\cal O}_{1},q}^{\mu\nu}\right],
\end{align}
where $q=c, d$ and $J_{{\cal O}_{1},q}, K_{{\cal O}_{1},q}$ are the effective four-quark currents
\begin{align}
J_{{\cal O}_{1},c}^{\mu} & =2i\ Q_c \lambda_c\ \bar{s}\gamma^{\nu}(1-\gamma_{5})\partial^{\mu}c\ \bar{u}\gamma_{\nu}(1-\gamma_{5})d,\nonumber\\
K_{{\cal O}_{1},c}^{\mu\nu} & =i\ Q_c \lambda_c\ \bar{s}\gamma^{\alpha}(1-\gamma_{5})\sigma^{\mu\nu}c\ \bar{u}\gamma_{\alpha}(1-\gamma_{5})d,\nonumber\\
J_{{\cal O}_{1},d}^{\mu} & =2i\ Q_d \lambda_d\ \bar{s}\gamma^{\nu}(1-\gamma_{5})c\ \bar{u}\gamma_{\nu}(1-\gamma_{5})\partial^{\mu}d,\nonumber\\
K_{{\cal O}_{1},d}^{\mu\nu} & =i\ Q_d \lambda_d\ \bar{s}\gamma^{\alpha}(1-\gamma_{5})c\ \bar{u}\gamma_{\alpha}(1-\gamma_{5})\sigma^{\mu\nu}d~.\label{eq:InitialRadiCurrents}
\end{align}
Here, $Q_q$ is the electric charge and $\lambda_q=\frac{m_{\Lambda_{c}}}{\bar m_{q}(m_{\Lambda_{c}}^{2}-
  m_{\Sigma}^{2})}$. For the case of ${\cal O}_2$ the corresponding operators can be obtained by just
exchanging the $u, s$ fields. Now the initial radiation amplitude induced by ${\cal O}_{1,2}$
can be expressed as:
\begin{align}
{\cal A}_{{\rm Initial}}^{{\cal O}_{i}}= -i \frac{G_{F}}{\sqrt{2}}V_{cs}V_{ud}^{*}\ C_i\ \varepsilon_{\mu}^{*}(k)
\left[\langle\Sigma(p)|J_{{\cal O}_{i}}^{\mu}(0)|\Lambda_{c}(q)\rangle+\langle\Sigma(p)|k_{\alpha}
K_{{\cal O}_{i}}^{\alpha\mu}(0)|\Lambda_{c}(q)\rangle\right],
\label{eq:InitialAmps}
\end{align}
where $i=1,2$, $J_{{\cal O}_{i}}^{\mu}=J_{{\cal O}_{i,c}}^{\mu}+J_{{\cal O}_{d}}^{\mu}$ and
$K_{{\cal O}_{i}}^{\mu\nu}=K_{{\cal O}_{i,c}}^{\mu\nu}+K_{{\cal O}_{d}}^{\mu\nu}$.  $k=q-p$ is the on-shell photon momentum. 

For the final quark radiation this effective Hamiltonian approach is not suitable. The reason is
that in our case the final baryon $\Sigma^+$ contains no heavy quark, and thus we cannot equate its
velocity with its constituent quarks, namely the momentum relation $p_{u,s}=(\bar m_{u,s}/m_{\Sigma})p_{\Sigma}$
cannot be used any more. The amplitude for the final quark radiation is calculated in the full theory.
It can be written as
\begin{align}
{\cal A}_{{\rm Final}}^{{\cal O}_i}= -i \frac{G_{F}}{\sqrt{2}}V_{cs}^{}V_{ud}^{*}\ C_i\ \varepsilon_{\mu}^{*}(k)\int d^4 x\  \langle\Sigma(p)|T\{ j^{\mu}(0){\cal O}_i(x)\}|\Lambda_{c}(q)\rangle,\label{eq:FinalAmps}
\end{align}
where $j^{\mu}=i Q_{u}\bar u \gamma^{\mu}u+i Q_{s}\bar s \gamma^{\mu}s$ is the quark electromagnetic current.
According to the Ward-identity, the matrix elements appearing in Eq.~(\ref{eq:InitialAmps}) and
Eq.~(\ref{eq:FinalAmps}) can be parameterized as
\begin{align}
\langle\Sigma(p)|{\cal J}_{{\cal O}_{i}}^{\mu}(0)|\Lambda_{c}(q)\rangle & = i\  \varepsilon^{*\mu}(k)\bar u_{\Sigma}
\left(a_{i, {\cal J}}^{+}+b_{i, {\cal J}}^{+} \gamma_5\right)\, \sigma^{\mu\nu}\frac{k_{\nu}}{m_{\Lambda_c}}u_{\Lambda_c}(q),
\label{eq:MatrixPars}
\end{align}
where ${\cal J}_{{\cal O}_{i}}^{\mu}=J_{{\cal O}_{i}}^{\mu}+ k_{\alpha}K_{{\cal O}_{i}}^{\alpha\mu}$ for the initial
radiation and  ${\cal J}_{{\cal O}_{i}}^{\mu}=\int d^4 x T\{ j^{\mu}{\cal O}_i(x)\}$ for the final radiation.
The amplitudes $a^+_{i, \cal J}, b^+_{i, \cal J}$ will be calculated using LCSR in the next section.

\section{Hadron level calculation in LCSR}
\label{sec:LCSRatHadronlevel}
Now we present the calculation of $\Lambda_{c}^{+}\to\Sigma^{+}\gamma$ decay width within  the LCSR approach.
To obtain the matrix elements given in Eq.~(\ref{eq:MatrixPars}), one has to define a suitable correlation
function and calculate it both at the hadron and the QCD level. Matching these two levels by the
quark-hadron duality enables us to extract the decay amplitudes. Here we define a two-point
correlation function as:
\begin{align}
\Pi_{{\cal O}_{i},{\cal J}}(p,q)= p^{\mu}\int d^{4}x\ e^{-iq\cdot x}\langle\Sigma(p)|T\{{{\cal J}_{\mu}^{{\cal O}_{i}}(0)
\bar{J}_{\Lambda_{c}}(x)}\}|0\rangle,
\label{eq:CorreFuncHLinitial}
\end{align}
where $\bar{J}_{\Lambda_{c}}$ is a current creating the $\Lambda_c$ baryon and its explicit form will be given
later. Here, we have contracted the correlation function with a momentum vector  $p^{\mu}$.
Without this contraction, the correlation function will have 12 independent structures
$\gamma_{\mu}, \gamma_{\mu}\gamma_{5}, \gamma_{\mu}\slashed q, \gamma_{\mu}\slashed q\gamma_{5}, p_{\mu},~p_{\mu}
\gamma_{5}, p_{\mu}\slashed q, p_{\mu}\slashed q\gamma_{5}, q_{\mu},~q_{\mu}\gamma_{5},~q_{\mu}\slashed q,~q_{\mu}
\slashed q\gamma_{5}, 1, \gamma_{5}, \slashed q$, and $\slashed q\gamma_{5}$.
However, in Eq.~(\ref{eq:MatrixPars}) there are only two independent amplitudes. Thus it will become
ambiguous which two of the 12 structures should be chosen to extract the two amplitudes.
Contracting the momentum vector  $p^{\mu}$ reduces the number of independent structures to four, namely
$1, \gamma_5, \slashed q, \slashed q \gamma_5$, which is still too much. This mismatch can be solved by
doubling the number of amplitudes, as will be explained next.

At the hadron level, this correlation function is calculated by inserting a complete set of states
between the two currents. The lowest single particle state should be explicitly kept while the
higher excited states will be attributed to the continuous spectrum. To match the four independent
structures of the correlation function with the number of decay amplitudes, we have to
introduce two extra amplitudes from the decay of the negative parity state $\Lambda_c({1/2}^-)$.
Similarly to Eq.~(\ref{eq:MatrixPars}), the corresponding amplitudes are 
\begin{align}
i\  \varepsilon^{*\mu}(k)\bar u_{\Sigma} \left(a_{i, {\cal J}}^{-}+b_{i, {\cal J}}^{-} \gamma_5\right)
\sigma_{\mu\nu}\frac{k^{\nu}}{m_{\Lambda_c}}(i\gamma_5)u_{\Lambda_c}(q).\label{eq:MatrixParsMinus}
\end{align}
Now we have four amplitudes $a_{i, {\cal J}}^{\pm}, b_{i, {\cal J}}^{\pm}$ mapping to the four structures
$1, \gamma_5, \slashed q, \slashed q \gamma_5$. Keeping both the two lowest states $\Lambda_c({1/2}^\pm)$
and attributing higher excited states to the continuous spectrum, we express the hadron level
correlation function of Eq.~(\ref{eq:CorreFuncHLinitial}) as
\begin{align}
\Pi_{{\cal O}_{i},{\cal J}}(p,q)_{\rm H} =& \frac{\lambda_{+}}{m_{\Lambda_{c}+}^{2}-q^{2}}\bar{u}_{\Sigma}(a_{i,{\cal J}}^{+}+b_{i,{\cal J}}^{+}\gamma_{5})\sigma_{\mu\nu}(\slashed q+m_{\Lambda_{c}+})\frac{p^{\mu}k^{\nu}}{m_{\Lambda_{c}+}}\nonumber\\
& +\frac{\lambda_{-}}{m_{\Lambda_{c}-}^{2}-q^{2}}\bar{u}_{\Sigma}(a_{i,{\cal J}}^{-}+b_{i,{\cal J}}^{-}\gamma_{5})\sigma_{\mu\nu}(\slashed q-m_{\Lambda_{c}-})\frac{p^{\mu}k^{\nu}}{m_{\Lambda_{c}-}}+\int_{s_{\rm th}}^{\infty}ds \frac{\rho_{{\cal O}_{i},{\cal J}}(s,p)}{s-q^2}~.
\label{eq:CorreFuncHLexpr}
\end{align}
The last term is the continuous spectrum contribution including all the states above the
$\Lambda_c({1/2}^-)$. $s_{\rm th}$ is the threshold parameter of this continuous spectrum and
should be larger than $m_{\Lambda_c -}^2$. $\lambda_{\pm}$ are the decay constants of the $\Lambda_c({1/2}^{\pm})$
which are defined as
\begin{align}
&\langle \Lambda_c({1/2}^+)(q)|\bar{J}_{\Lambda_{c}}(0)|0\rangle=\bar u_{\Lambda_c}(q)\lambda_{+},\nonumber\\
&\langle \Lambda_c({1/2}^-)(q)|\bar{J}_{\Lambda_{c}}(0)|0\rangle=\bar u_{\Lambda_c}(q)(i\gamma_5)\lambda_{-}.
\end{align}

The same correlation function should also be calculated at the QCD level, which can be 
expressed as a dispersion integral:
\begin{align}
\Pi_{{\cal O}_{i},{\cal J}}(p,q)_{\rm QCD}
=\frac{1}{2\pi i}\int_{m_c^2}^{\infty} ds \frac{{\rm Disc}\ \Pi_{{\cal O}_{i},{\cal J}}(p,s)_{\rm QCD}}{s-q^2}.
\label{dispersiveQCD}
\end{align}
The discontinuity part can be parameterized as:
\begin{align}
{\rm Disc}\ \Pi_{{\cal O}_{i},{\cal J}}(s,p)_{\rm QCD}=F^{(1)}_{{\cal O}_{i},{\cal J}}\ \slashed q
\gamma_5+F^{(2)}_{{\cal O}_{i},{\cal J}}\ \slashed q+F^{(3)}_{{\cal O}_{i},{\cal J}}\ \gamma_5+F^{(4)}_{{\cal O}_{i},{\cal J}}.
\end{align}
In principle, the correlation function calculated at the hadron and the QCD level should be equivalent.
According to the quark-hadron duality, the continuous spectrum contribution in Eq.~(\ref{eq:CorreFuncHLexpr})
is canceled by the corresponding QCD level dispersion integral in the region $s_{\rm th}<s<\infty$.
Furthermore, since the QCD level calculation can only be explicitly performed using a light-cone expansion
(LCE), one has to perform a Borel transformation of the correlation function at both levels to improve
the LCE convergence. Finally one can extract the amplitudes as
\begin{align}
a_{i,{\cal J}}^{+} & =\frac{1}{\pi}\int_{m_{c}^{2}}^{s_{{\rm th}}}ds\ e^{\frac{m_{\Lambda_{c}+}^{2}-s}{T_{2}}}\frac{m_{\Lambda_{c}+}\big[m_{\Lambda_{c}-}F_{{\cal O}_{i},{\cal J}}^{(2)}+F_{{\cal O}_{i},{\cal J}}^{(4)}\big]}{\lambda_{+}(m_{\Lambda_{c}+}+m_{\Lambda_{c}-})(m_{\Lambda_{c}+}-m_{\Sigma})^{2}},\nonumber\\
b_{i,{\cal J}}^{+} & =\frac{1}{\pi}\int_{m_{c}^{2}}^{s_{{\rm th}}}ds\ e^{\frac{m_{\Lambda_{c}+}^{2}-s}{T_{2}}}\frac{m_{\Lambda_{c}+}\big[F_{{\cal O}_{i},{\cal J}}^{(3)}-m_{\Lambda_{c}-}F_{{\cal O}_{i},{\cal J}}^{(1)}\big]}{\lambda_{+}(m_{\Lambda_{c}+}+m_{\Lambda_{c}-})(m_{\Lambda_{c}+}+m_{\Sigma})^{2}},\label{eq:ExtractAmps}
\end{align}
where $T_2$ is the Borel parameter which will be determined during the numerical calculation.
Here, $a_{i,{\cal J}}^{-}, b_{i,{\cal J}}^{-}$ are not shown since we only care about the decay amplitudes
of the $\Lambda_c({1/2}^{+})$. The coefficients $F_{{\cal O}_{i},{\cal J}}^{(n)}$ will be explicitly calculated
by the LCE at the QCD level.

\section{QCD level calculation in LCSR}
\label{sec:LCSRatQCDlevel}

In this section, we will use light-cone expansion to calculate the correlation function
defined in Eq.~(\ref{eq:CorreFuncHLinitial}), and extract the coefficients $F_{{\cal O}_{i},{\cal J}}^{(n)}$.
Now the correlation function reads 
\begin{align}
&\Pi_{{\cal O}_{i},{\cal J}}(p,q)_{\rm QCD} = p^{\mu}\int d^{4}x\ e^{-iq\cdot x}\langle\Sigma(p)|
T\{{{\cal J}_{\mu}^{{\cal O}_{i}}(0)\bar{J}_{\Lambda_{c}}(x)}\}|0\rangle,\nonumber\\
&{\rm with}~~~~\bar{J}_{\Lambda_{Q}} =-\epsilon_{abc}\bar{Q}_{c}(\bar{d}_{b}C\gamma_{5}\bar{u}_{a}^{T}),
\label{eq:CorreFuncQCDLinitial}
\end{align}
where $a,b,c$ are color indices. Here, $q^2\ll 0$ is taken in the deep Euclidean region to realize
the light-cone expansion.  Let us take $i=1$ and ${\cal J}=J$ as an example to illustrate the
detailed calculation for the $c$ quark radiation. 

\begin{figure}
\begin{center}
\includegraphics[width=0.85\columnwidth]{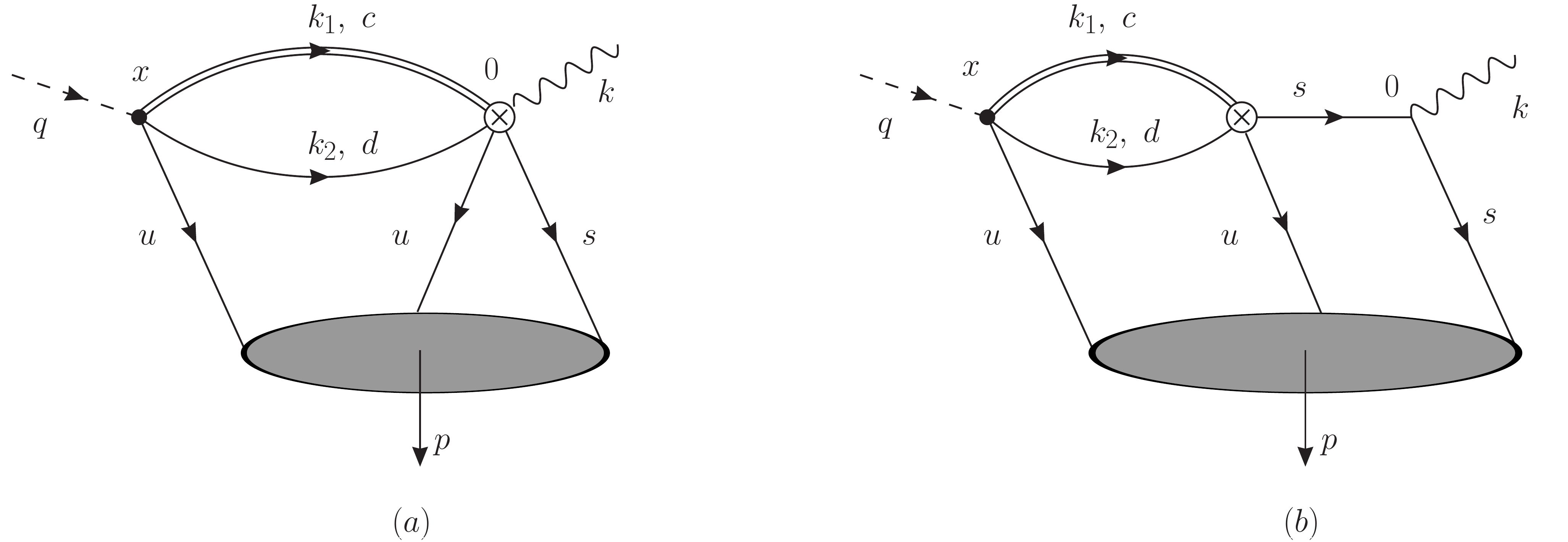} 
\caption{Diagrams for the QCD level correlation function in Eq.~(\ref{eq:CorreFuncQCDLinitial}). (a)
is the initial quark radiation where the white crossed dot denotes the effective four-quark
currents in Eq.~(\ref{eq:InitialRadiCurrents}). (b) is the final $s$ quark radiation where the
white crossed dot denotes the current $j^{\mu}$.  The black dot denotes the $\Lambda_c$ current.
The grey ellipse represents the LCDAs of the $\Sigma^+$.}
\label{fig:LcToSigmaInitial} 
\end{center}
\end{figure}

At leading order the corresponding correlation function becomes
\begin{align}
\Pi_{{\cal O}_{i},{\cal J}}^{c}(p,q)_{\rm QCD} =& -2i\ Q_c \lambda_c \epsilon_{abc}\  p_{\mu}\int d^{4}x\
e^{-iq\cdot x}[\gamma^{\nu}(1-\gamma_{5})\partial_{w}^{\mu}S_{c}(w,x)]_{ig}\nonumber\\
& \times[\gamma_{\nu}(1-\gamma_{5})S_{d}(0,x)C\gamma_{5}]_{kn}\langle\Sigma(p)|\bar{s}_{c}^{i}(0)
\bar{u}_{b}^{k}(0)\bar{u}_{a}^{n}(x)|0\rangle,
\label{eq:QCDcorreO1J}
\end{align}
where $S_q(x,y)$ is the free propagator of the quark $q$. Fig.~\ref{fig:LcToSigmaInitial}(a)
shows the corresponding Feynman diagram, where the black dot at coordinate $x$ denotes the $\Lambda_c$ current 
 and
the white crossed dot at coordinate $0$ denotes the effective current $J_{{\cal O}_{1},c}^{\mu}$.
The last matrix element in Eq.~(\ref{eq:QCDcorreO1J}) is represented by the grey ellipse in
Fig.~\ref{fig:LcToSigmaInitial}, which can be parameterized by three leading twist LCDAs of
the $\Sigma^+$\cite{Lepage:1980fj,Efremov:1979qk,Chernyak:1983ej,Krankl:2011gch,RQCD:2019hps}
\begin{align}
\langle\Sigma(p)|\bar{s}_{c}^{i}(0)\bar{u}_{b}^{k}(0)\bar{u}_{a}^{n}(x)|0\rangle =& -\frac{1}{4}\epsilon_{abc}\int du_{1}du_{2}du_{3}\delta(1-u_{1}-u_{2}-u_{3})\ e^{iu_{1}p\cdot x}\nonumber\\
 &\times\Big\{[\bar{u}^{B}\gamma_{5}]_i [C \slashed {\tilde n}]_{kn}V^{B}(u_{1},u_{2},u_{3})+[\bar{u}^{B}]_i [C\gamma_{5} \slashed {\tilde n}]_{kn}A^B(u_{1},u_{2},u_{3})\nonumber\\
 &+\ i\tilde{n}^{\alpha}g_{\perp}^{\beta\rho}[C\sigma_{\beta\alpha}]_{kn}[\bar{u}^{B}\gamma_{\rho}\gamma_{5}]_{i}T^{B}(u_{1},u_{2},u_{3})\Big\},
\end{align}
where $g_{\mu \nu}^{\perp}=g_{\mu \nu}-(\tilde{n}_{\mu} n_{\nu}+\tilde{n}_{\nu} n_{\mu})/(\tilde{n} \cdot n)$
with $\tilde{n}_{\mu}=p_{\mu}-(m_{\Sigma}^{2}/2p \cdot n) n_{\mu}$ and $n$ is a light-cone vector.
$\bar{u}^{B}=\bar{u}_{\Sigma}\slashed n\slashed p/2m_{\Sigma}$ with $u_{\Sigma}$ the Dirac spinor of the
$\Sigma^+$ baryon.  The coordinate $x$ of the quark field is parallel to  $n$, $x=(x\cdot p/m_{\Sigma})n$,
where we have used $p=m_{\Sigma} v$, $v=(n+\bar n)/2$ and $n\cdot\bar n=2$. In the chiral limit $m_u=m_d=0$
the contribution of $V^B$ and $A^B$ to the correlation function vanishes. The explicit form of
$T^B$ of the $\Sigma^+$ baryon now reads
\begin{align}
T^B(u_{1},u_{2},u_{3})=120 u_{1} u_{2} u_{3}\left(\pi_{00}^{B} \mathcal{P}_{00}
+ \pi_{11}^{B} \mathcal{P}_{11}+\ldots\right),
\end{align}
where the $\mathcal{P}_{ij}$ are polynomials, $\mathcal{P}_{00}=1,~~ \mathcal{P}_{11}=7\left(u_{1}-2 u_{3}+u_{2}
\right)$ \cite{RQCD:2019hps}.
$\pi_{00}^B$ and $\pi_{11}^B$ are the shape parameters which encode all non-perturbative information of
the baryon. The ellipsis denotes terms of higher power polynomials, which are suppressed and omitted here.

Using the $\Sigma^+$ LCDAs given above, we can express the correlation function in
Fig.~\ref{fig:LcToSigmaInitial}(a) as
\begin{align}
\Pi_{{\cal O}_{i},{\cal J}}^{c}(p,q)_{\rm QCD} =& -3i Q_c \lambda_c \frac{1}{2m_{\Sigma}}\int du_{1}du_{2}\int d^{4}x\int\frac{d^{4}k_{1}}{(2\pi)^{4}}\frac{d^{4}k_{2}}{(2\pi)^{4}}\ e^{-i(q-k_{1}-k_{2}-u_{1}p)\cdot x}\nonumber\\
 & \times T^{B}(u_{1},u_{2},1-u_{1}-u_{2})\frac{1}{k_{1}^{2}-m_{c}^{2}}\frac{1}{k_{2}^{2}}\ p\cdot k_1\nonumber\\
 & \times \left[p^{\alpha}g^{\rho\beta}+(\frac{1}{2}m_{\Sigma}g^{\rho\alpha}-\frac{1}{m_{\Sigma}}p^{\rho}p^{\alpha})n^{\beta}+\frac{1}{2}p^{\alpha}n^{\beta}n^{\rho}\right]n^{\kappa}\nonumber\\
 & \times\bar{u}_{\Sigma}\gamma_{\kappa}\slashed p\gamma_{\rho}\gamma_{5}\gamma^{\nu}(1-\gamma_{5})(\slashed k_{1}+m_{c})\ {\rm tr}[\gamma_{\nu}(1-\gamma_{5})\slashed k_{2}\sigma_{\alpha\beta}].
 \end{align}
Here, we have defined $\tilde{T}^{B}(u_{1},u_{2})=T^{B}(u_{1},u_{2},1-u_{1}-u_{2})$. Note that since
$n=(m_{\Sigma}/x\cdot p)x$, we can use the following trick to remove the $x$ in the denominator:
\begin{align}
 & \int du_{1}du_{2}\int d^{4}x\ e^{-i(q-k_{1}-k_{2}-u_{1}p)\cdot x}\ \tilde{T}^{B}(u_{1},u_{2})n_{\kappa}\ \cdots\nonumber\\
=\ & m_{\Sigma}\frac{\partial}{\partial q^{\kappa}}\int du_{1}du_{2}\int d^{4}x\ e^{-i(q-k_{1}-k_{2}-u_{1}p)\cdot x}\ \tilde{T}_{(1)}^{B}(u_{1},u_{2})\ \cdots,\label{eq:replaceN}
\end{align}
where the ellipses represents all the terms independent of $u_1, u_2$, and
\begin{align}
 \tilde{T}_{(i)}^{B}(u_{1},u_{2})=\int_{0}^{u_{1}}dt\ \tilde{T}_{(i-1)}^{B}(t,u_{2})~~~~~{\rm with}~~~\tilde{T}_{(0)}^{B}(t,u_{2})=\tilde{T}^{B}(t,u_{2}).
\end{align}
From Eq.~(\ref{eq:replaceN}), it follows that for each $n_{\kappa}$ one can equivalently replace it
with an operator $\hat{n}_{\kappa}=m_{\Sigma}\ \partial/\partial q^{\kappa}$ and simultaneously
replace $\tilde{T}^{B}$ with $\tilde{T}_{(1)}^{B}$. Therefore, the correlation function takes the form
\begin{align}
\Pi_{{\cal O}_{i},{\cal J}}^{c}(p,q)_{\rm QCD}=& -3i Q_c \lambda_c\frac{1}{2m_{\Sigma}}\int du_{1}du_{2}\
{\cal N}[\hat n, T_{(i)}^B]^{\alpha\beta\rho\kappa}\int\frac{d^{4}k_{1}}{(2\pi)^{4}}\frac{d^{4}k_{2}}{(2\pi)^{4}}
\nonumber\\
&\times(2\pi)^{4}\delta^{4}(q-u_{1}p-k_{1}-k_{2})\frac{1}{k_{1}^{2}-m_{c}^{2}}\frac{1}{k_{2}^{2}}\ p\cdot k_{1}
\nonumber\\
& \times\bar{u}_{\Sigma}\gamma_{\kappa}\slashed p\gamma_{\rho}\gamma_{5}\gamma^{\nu}(1-\gamma_{5})(\slashed k_{1}
+m_{c})\ {\rm tr}[\gamma_{\nu}(1-\gamma_{5})\slashed k_{2}\sigma_{\alpha\beta}],
\end{align}
where the operator $\cal N$ is defined as
\begin{align}
&{\cal N}[\hat n, T_{(i)}^B]^{\alpha\beta\rho\kappa}\nonumber\\
=&\  \hat{n}^{\kappa}\left[p^{\alpha}g^{\rho\beta}\tilde{T}_{(1)}^{B}(u_{1},u_{2})+\left(\frac{1}{2}m_{\Sigma}g^{\rho\alpha}-\frac{1}{m_{\Sigma}}p^{\rho}p^{\alpha}\right)\hat{n}^{\beta}\tilde{T}_{(2)}^{B}(u_{1},u_{2})+\frac{1}{2}p^{\alpha}\hat{n}^{\beta}\hat{n}^{\rho}\tilde{T}_{(3)}^{B}(u_{1},u_{2})\right].
\end{align}

Now we have to express the QCD level correlation function as a dispersive integral. The discontinuity
part can be extracted from the cutting rules:
\begin{align}
{\rm Disc}\ \Pi_{{\cal O}_{i},{\cal J}}^{c}(p,q)_{\rm QCD}=& -3i Q_c \lambda_c\frac{(2\pi)^2}{2m_{\Sigma}}{\cal N}[\hat n, T_{(i)}^B]^{\alpha\beta\rho\kappa}\int du_{1}du_{2}\int d\Phi_2[(q-u_1 p)^2] \ p\cdot k_1\nonumber\\
 & \times\bar{u}_{\Sigma}\gamma_{\kappa}\slashed p\gamma_{\rho}\gamma_{5}\gamma^{\nu}(1-\gamma_{5})(\slashed k_{1}+m_{c})\ {\rm tr}[\gamma_{\nu}(1-\gamma_{5})\slashed k_{2}\sigma_{\alpha\beta}],\label{eq:DiscPiQCD}
\end{align}
where 
\begin{align}
d\Phi_2[(q-u_1 p)^2]=\int\frac{d^{3}k_{1}}{(2\pi)^{3}}\frac{1}{2E_{k_{1}}}\frac{d^{3}k_{1}}{(2\pi)^{3}}\frac{1}{2E_{k_{1}}}\delta^{4}(q-u_1 p-k_{1}-k_{2})
\end{align}
is the two-body phase space integration, which corresponds to cutting off the $c,d$ quark loop
in Fig.~\ref{fig:LcToSigmaInitial}(a). Further, $\Pi_{{\cal O}_{2},{\cal J}}^{c}(p,q)_{\rm QCD}=
-\Pi_{{\cal O}_{1},{\cal J}}^{c}(p,q)_{\rm QCD}$ so that we only have to calculate the amplitudes induced
by ${\cal O}_1$. The integration in Eq.~(\ref{eq:DiscPiQCD}) is involved but straightforward,
so we will not present further calculational details here.

For the case of the final quark radiation, the corresponding diagram is shown in
Fig.~\ref{fig:LcToSigmaInitial}(b), where we take the $s$ quark radiation as an example. The calculation
for this diagram is similar to Fig.~\ref{fig:LcToSigmaInitial}(a) and the only difference is that now
we have an extra $s$ quark propagator:
\begin{align}
\frac{1}{(q-(u_{1}+u_{2})p)^{2}}=\frac{1}{u_{3}(s-(u_{1}+u_{2})m_{\Sigma}^{2})}.\label{eq:extraProp}
\end{align}
It should be mentioned that for the final quark radiation
${\cal J}_{\mu}^{{\cal O}_{i}}(0)$ is an composite operator of $j_{q^{\prime}\mu}$ and ${\cal O}_i$, so
that the hadron level correlation function in Eq.~(\ref{eq:CorreFuncHLinitial}) is actually induced
by three operators. However, since we only insert a complete set of states between
${\cal J}_{\mu}^{{\cal O}_{i}}(0)$ and $\bar{J}_{\Lambda_{c}}(x)$, the composite operator
${\cal J}_{\mu}^{{\cal O}_{i}}(0)$ is not disconnected. Therefore, at the QCD level when extracting the
discontinuity part, we only have to cut off the  $c,d$ quark loop in Fig.~\ref{fig:LcToSigmaInitial}(b)
and keep the $s$ quark propagator unchanged.

\section{Numerical Results}\label{sec:numericalresult}
\label{sec:numericalResult}

We first give the input parameters.
We use the $\overline{\rm MS}$ masses for the quarks, $m_c(\mu)=1.27$~GeV and $m_s(\mu)=0.103$~GeV with
$\mu= 1.27$ GeV \cite{ParticleDataGroup:2020ssz}. The masses of $u,d$ quarks are omitted. The composite
masses of  the $c,d$ quarks are taken as $\bar m_c=1.6$~GeV and $\bar m_d=0.32$~GeV \cite{Cheng:1994kp}.
The masses of the baryons are $m_{\Sigma}=1.19$~GeV, $m_{\Lambda_c+}=2.286$~GeV and
$m_{\Lambda_c-}=2.6$~GeV \cite{ParticleDataGroup:2020ssz}. The decay constant of the
$\Lambda_c({1/2}^+)$ is taken as $\lambda_+=0.01\pm 0.001$ \cite{Zhao:2020mod}. From
Eq.~(\ref{eq:ExtractAmps}), it can be seen that the amplitudes are proportional to the inverse of $\lambda_+$
so that its uncertainty may affect the result a lot. Therefore, we will include the uncertainty of
$\lambda_+$ when evaluating the uncertainty of the decay amplitudes.  
The shape parameters of the $\Sigma^+$ LCDAs are taken from a lattice calculation with $N_f=2+1$ and
vanishing lattice spacing limit $a\to 0$ :  $\pi_{00}^B=5.14\times 10^{-3}$~GeV$^2$ and
$\pi_{11}^B=-0.09\times 10^{-3}$~GeV$^2$\cite{RQCD:2019hps}.

Further, the LCSR contains two kinds of extra parameters, namely the threshold parameter $s_{\rm th}$ and
the Borel parameter $T_2$. 
The threshold parameter should in principle be process independent and only related to the corresponding
hadron state. Here, $s_{\rm th}$ is taken from a QCD sum rules study on the decay constant of
the $\Lambda_c$ \cite{Zhao:2020mod}: $s_{\rm th}=2.85^2$ GeV$^2$. Generally, the sum rules results are sensitive to the
threshold parameter, thus here we consider a small uncertainty $\pm 0.5$ GeV$^2$ near this value 
to evaluate the uncertainty from the threshold parameter on the decay amplitudes. 

\begin{figure}
\begin{center}
\includegraphics[width=0.45\columnwidth]{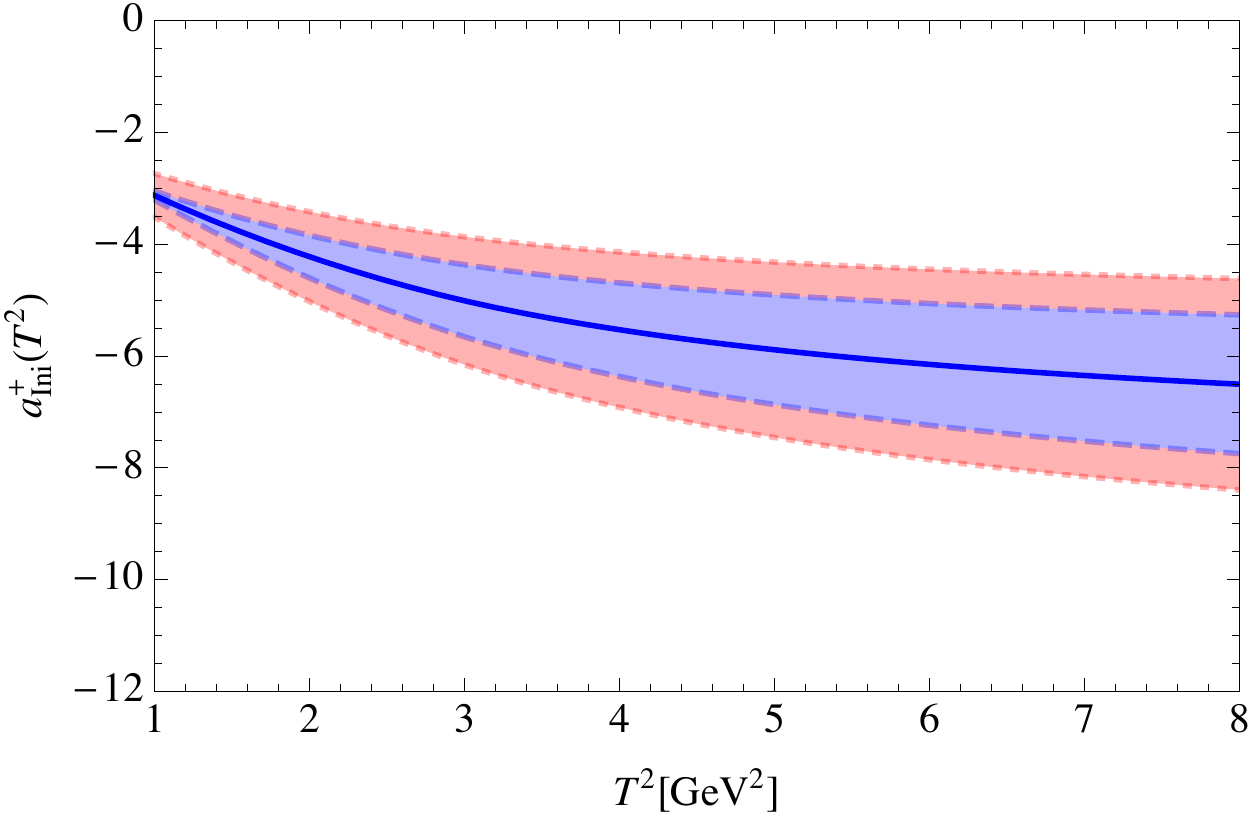} 
\includegraphics[width=0.45\columnwidth]{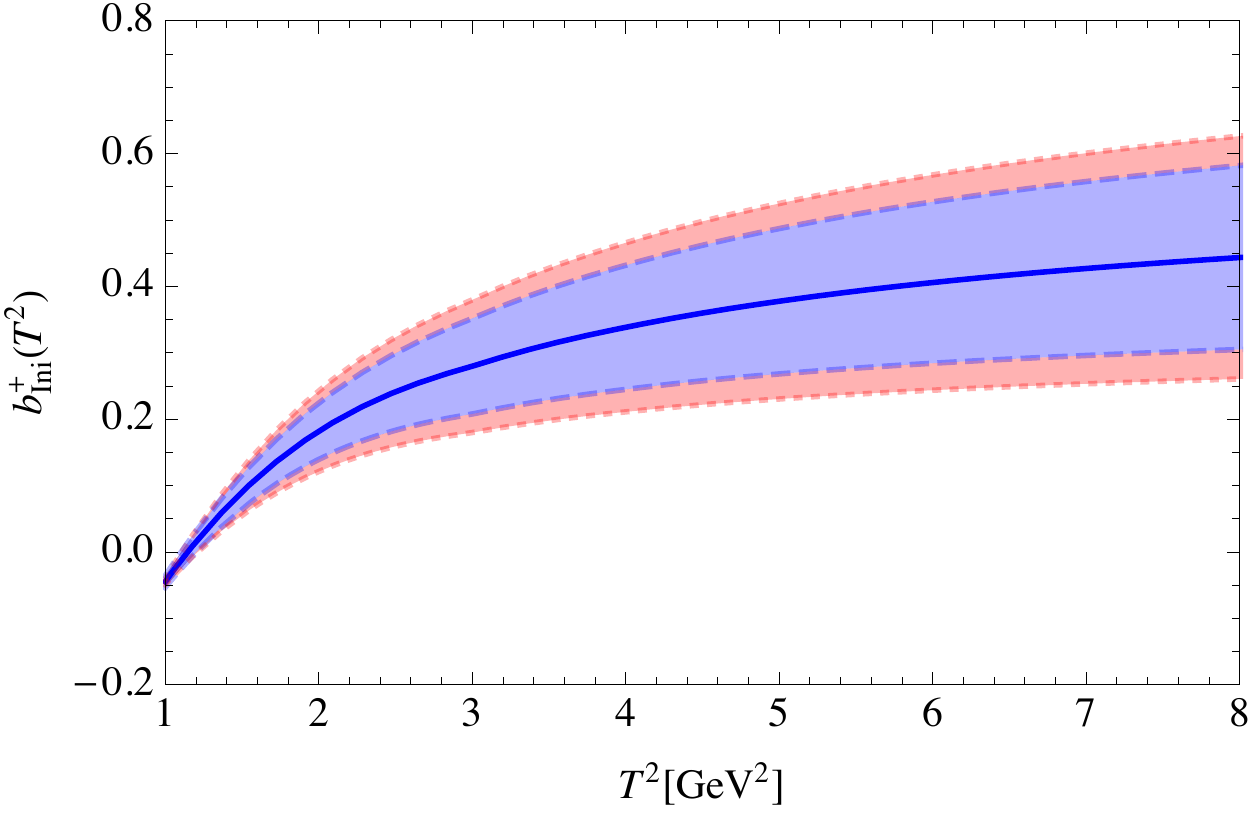} 
\includegraphics[width=0.45\columnwidth]{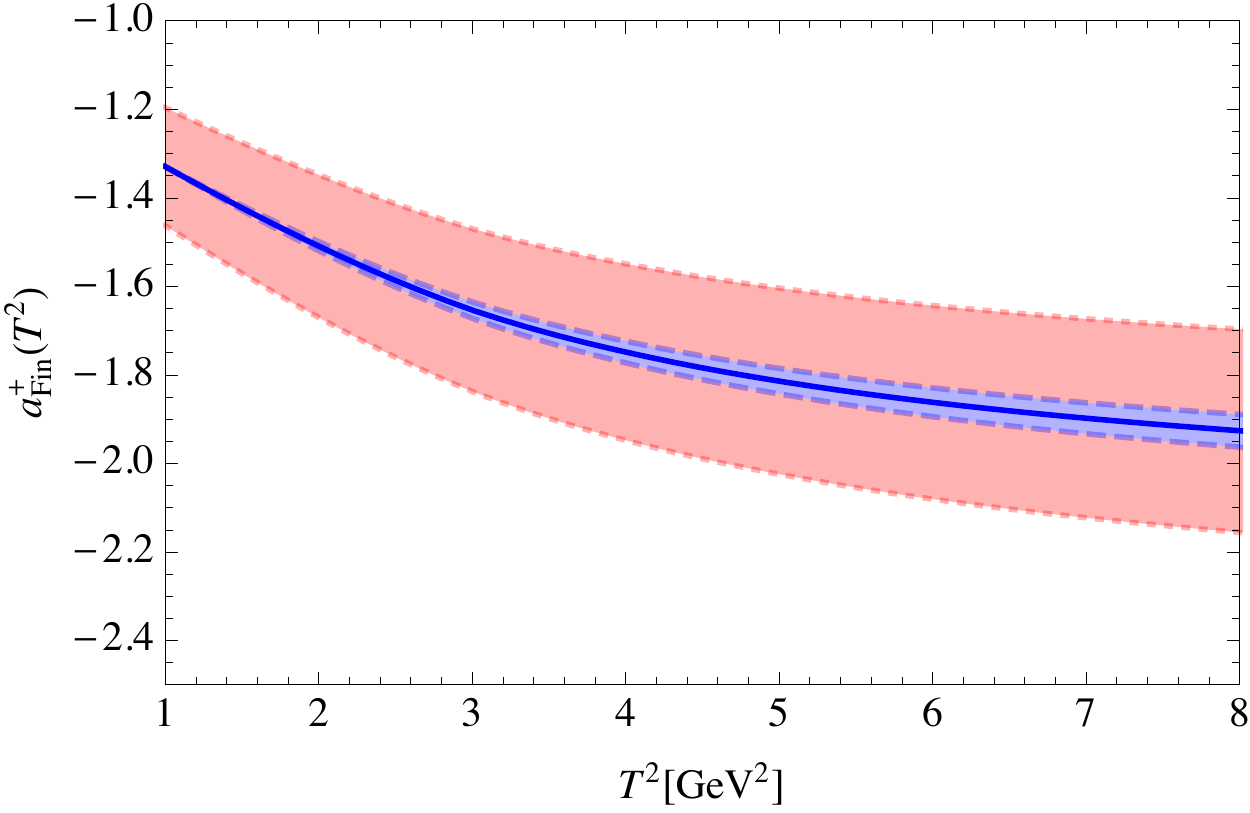} 
\includegraphics[width=0.45\columnwidth]{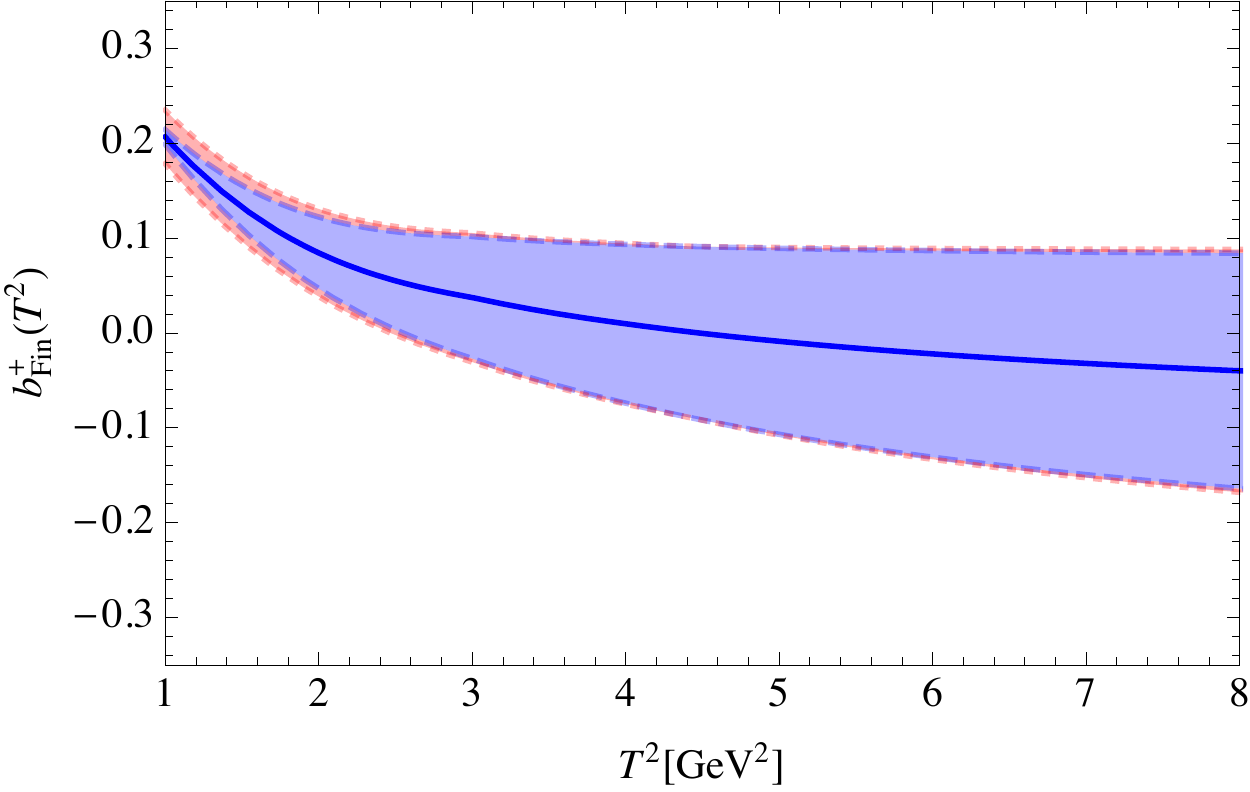} 
\caption{Decay amplitudes $a_{\cal J}^+$ and $b_{\cal J}^+$ (in unit $10^{-3}$~GeV$^2$) as functions of the
  Borel parameter $T^2$. In each diagram, the blue band denotes the error from the uncertainty of
  the threshold $s_{\rm th}=2.85^2\pm 0.5$ GeV$^2$. 
  The upper and lower red bands denote the error from
  the uncertainty of $\lambda_+$.}
\label{fig:aPlusbPlusVST2} 
\end{center}
\end{figure}

Generally, the Borel parameter $T_2$ is chosen to satisfy three requirements. First, $T_2$ cannot be
too large so that the continuous spectrum  contribution is suppressed. Second, $T_2$ must be large
enough to ensure the light-cone expansion to  convergence. Finally, the result must be stable in a
window of $T_2$. The first and the second requirement can determine the upper and lower bound of the
$T_2$ window, respectively. Fig.~\ref{fig:aPlusbPlusVST2} shows the amplitudes $a_{i, {\cal J}}^{+}$ and
$b_{i, {\cal J}}^{+}$ as functions of $T_2$. To determine the upper bound, we require that the
pole contribution must be larger than the continuous spectrum contribution, namely:
\begin{align}
\displaystyle\frac{\displaystyle\int_{m_{c}^{2}}^{s_{\rm th}}ds\  e^{-s/T^{2}}{\rm Disc}\ \Pi_{{\cal O}_{i},{\cal J}}(p,s)_{\rm QCD}}
{\displaystyle\int_{m_{c}^{2}}^{\infty}ds\  e^{-s/T^{2}}{\rm Disc}\ \Pi_{{\cal O}_{i},{\cal J}}(p,s)_{\rm QCD}}>0.5.
\end{align}

The numerator is the pole contribution, which represents the integral on the right-hand side of
Eq.~(\ref{eq:ExtractAmps}). The denominator is the same integral but the upper limit of $s$ is
extended to infinity, which contains both pole and continuous spectrum contributions. 
Note that although the value for  this fraction is derived from experience, as long as the
third requirement for stability is satisfied, 
the result will be insensitive to this fraction, and its uncertainty can be attributed to choosing the window of $T_2$.

On the other hand, in principle,  the lower bound of the $T_2$ is determined by the ratio between
the contribution from the leading order and next-to-leading order QCD corrections to the perturbative
kernel. However, in this work only the leading order contribution is considered so that this method
cannot be used.  Following our previous work~\cite{Shi:2022kfa}, to get  the window of $T_2$,
we can set the center value of $T_2$ as its upper bound, and find a range $\pm 1$ GeV$^2$ around
this center value.  The amplitudes and the corresponding errors from the
uncertainties of $s_{\rm th}$, $T_2$ and $\lambda_+$ are listed in Table~\ref{Tab:decayAmps}.
Note that the center value of the $T_2$ is already in a relatively stable region as shown in
Fig.~\ref{fig:aPlusbPlusVST2}, thus the procedure given above is sufficient for determining
the errors of the amplitudes. Generally, the Borel parameters are close to the corresponding mass square of hadrons. From Table~\ref{Tab:decayAmps}, the $T_2$ s for initial-state radiation are close to $m_{\Lambda_c}^2$ which is as expected. However, the $T_2$ s for final-state radiation are much smaller. The reason is that in Fig.~\ref{fig:LcToSigmaInitial}(b) the extra propagator as shown in Eq.~(\ref{eq:extraProp}) provides a lighter mass scale $m_{\Sigma}$. Now the $s$ dominates around $m_{\Sigma}^2$, which reduces the optimal value of $T_2$.
\begin{table}
\caption{Decay amplitudes $a_{\cal J}^+$ and $b_{\cal J}^+$ (in unit $10^{-3}$~GeV$^2$) and the
    corresponding Borel parameters (GeV$^2$) for the initial- and the final-state radiation.}
\label{Tab:decayAmps}
\begin{tabular}{|cc|cc|cc|cc|}
\hline 
$a_{{\rm Ini}}^{+}$ & $T_{2}$ & $b_{{\rm Ini}}^{+}$ & $T_{2}$ & $a_{{\rm Fin}}^{+}$ & $T_{2}$ & $b_{{\rm Fin}}^{+}$ & $T_{2}$\tabularnewline
\hline 
$-6.03\pm 1.22$~~ & $5.5\pm1.0$ & $0.37\pm 0.11$~~ & $4.7\pm1.0$ & $-1.56\pm 0.17$~~  & $2.3\pm 0.5$  & $0.13\pm 0.05$~~ & $1.45\pm 0.5$  \tabularnewline
\hline 
\end{tabular}
\end{table}

Using the amplitudes given in Table~\ref{Tab:decayAmps}, we can obtain the decay width of the
$\Lambda_{c}^{+}\to\Sigma^{+}\gamma$ from the formula 
\begin{align}
\Gamma\left(\Lambda_{c}^{+}\to\Sigma^{+}\gamma\right)=\frac{1}{8 \pi \
m_{\Lambda_{c}+}^{2}}\left(\frac{m_{\Lambda_{c}+}^{2}-m_{\Sigma}^{2}}{m_{\Lambda_{c}+}}\right)^{3}\frac{G_F^2}{2}
|V_{cs}V_{ud}|^2 (C_1-C_2)^2 \left(|a|^{2}+|b|^{2}\right)
\end{align}
with $a=a_{{\rm Ini}}^{+}+a_{{\rm Fin}}^{+}$ and $b=b_{{\rm Ini}}^{+}+b_{{\rm Fin}}^{+}$. The Wilson coefficients are
taken as $C_1=1.22$ and $C_2=-0.43$ at $\mu=m_c$ \cite{Li:2012cfa}. The CKM matrix elements are 
$|V_{cs}|=0.975$ and $|V_{ud}|=0.973$ \cite{ParticleDataGroup:2020ssz}. Using the $\Lambda_c^+$ lifetime
$\tau(\Lambda_c^+)=2.01\times 10^{-13}$ s \cite{ParticleDataGroup:2020ssz}, we can obtain the branching
fraction
\begin{align}
{\cal B}(\Lambda_{c}^{+}\to\Sigma^{+}\gamma)=1.03\pm 0.36 \times 10^{-4},
\end{align}
which is below the experimental upper limit given recently by the Belle Collaboration \cite{Belle:2022raw}:
\begin{align}
{\cal B}_{\rm expr}(\Lambda_{c}^{+}\to\Sigma^{+}\gamma)<2.6 \times 10^{-4}.
\end{align}

Table~\ref{Tab:ComparedecayBr} gives a comparison of the $\Lambda_{c}^{+}\to\Sigma^{+}\gamma$ branching
fraction from this work, the result from the Belle Collaboration, the modified nonrelativistic
quark model (NRQM) \cite{Kamal:1983zt}, the constituent quark model (CQM) \cite{Uppal:1992cc} and
the effective Hamiltonian approach (EHA) \cite{Cheng:1994kp}. The branching fraction from the CQM is
slightly larger than the experimental upper limit, while the branching fractions from other
theoretical methods are nearly one order smaller than the upper limit. Our result is between
these theoretical predictions and the experimental upper limit. Due to the limitation on the
data sample and resolution, an extremely small branching fraction is difficult to be measured.
However, the relatively larger branching fraction predicted in this work is more likely to be tested
by future experiments.
\begin{table}
  \caption{Comparison of the branching fraction ${\cal B}(\Lambda_{c}^{+}\to\Sigma^{+}\gamma)$ from this work with those
    from the literature and the Belle experiment.}
\label{Tab:ComparedecayBr}
\begin{tabular}{|c|c|}
\hline 
Method & ${\cal B}(\Lambda_{c}^{+}\to\Sigma^{+}\gamma)$ \tabularnewline
\hline 
This Work & $1.03\pm0.36\times10^{-4}$\tabularnewline
NRQM \cite{Kamal:1983zt} & $3.2\times10^{-5}$\tabularnewline
CQM \cite{Uppal:1992cc} & $2.8\pm0.6\times10^{-4}$\tabularnewline
EHA \cite{Cheng:1994kp} & $4.9\times10^{-5}$\tabularnewline
\hline
Exp. \cite{Belle:2022raw} & $<2.6\times10^{-4}$\tabularnewline
\hline
\end{tabular}
\end{table}

\section{Conclusion}
\label{sec:conclusion}
We have calculated the decay width of  $\Lambda_{c}^{+}\to\Sigma^{+}\gamma$ using
light-cone sum rules. For the initial quark radiation we constructed an effective Hamiltonian to simplify
the calculation, where the internal quark line shrinks to a point. The final quark radiation is
studied utilizing the full theory. The leading twist light-cone distribution amplitudes of the
$\Sigma^+$ serve as the non-perturbative input for the sum rule calculation, and the perturbative
kernel is calculated at leading order. The branching fraction we obtain is
${\cal B}(\Lambda_{c}^{+}\to\Sigma^{+}\gamma)=1.03\pm 0.36 \times 10^{-4}$, which is between previous
theoretical predictions and the experimental upper limit. Considering the data sample and resolution
of the experiment, we believe that our prediction can be tested in the near future.

\section*{Acknowledgements}
The authors are grateful to Ji-Bo He, Chengping Shen, Wei Wang, Zhen-Xing Zhao and Chien-Yeah Seng for useful discussions.
This work is supported in part by the NSFC under Grant No.12147147 and the Deutsche Forschungsgemeinschaft (DFG, German Research
Foundation) through the funds provided to the Sino-German Collaborative
Research Center TRR110 ``Symmetries and the Emergence of Structure in QCD''
(NSFC Grant  No. 12070131001, DFG Project-ID 196253076 - TRR 110). 
The work of UGM was supported in part by the Chinese
Academy of Sciences (CAS) President's International
Fellowship Initiative (PIFI) (Grant No. 2018DM0034)
and by VolkswagenStiftung (Grant No. 93562).

\end{document}